\documentclass[reprint,superscriptaddress,twocolumn, aip, apl]{revtex4-1}

\usepackage{graphicx,psfrag}
\usepackage{epstopdf}
\usepackage{amsmath} 
\usepackage{bm}
\usepackage{amssymb}
\usepackage{subfigure}
\usepackage{color}
\usepackage[pagewise]{lineno}

\begin{document}

\title{\large \bf Total-internal-reflection elastic metasurfaces: design and application to structural vibration isolation}

\author{Hongfei Zhu}
\email{zhu307@purdue.edu}
\affiliation{Ray W. Herrick Laboratories, School of Mechanical Engineering, Purdue University, West Lafayette, Indiana 47907, USA}
\author {Timothy F. Walsh}
\email{tfwalsh@sandia.gov; Sandia National Laboratories is a multimission laboratory managed and operated by National Technology and Engineering Solutions of Sandia, LLC., a wholly owned subsidiary of Honeywell International, Inc., for the U.S. Department of Energy’s National Nuclear Security Administration. With main facilities in Albuquerque, N.M., and Livermore, C.A., Sandia has major R\&D responsibilities in national security, energy and environmental technologies, and economic competitiveness.}
\affiliation{Sandia National Laboratory, Albuquerque, New Mexico 87185, USA}
\author {Fabio Semperlotti}
\email{To whom correspondence should be addressed: fsemperl@purdue.edu}
\affiliation{Ray W. Herrick Laboratories, School of Mechanical Engineering, Purdue University, West Lafayette, Indiana 47907, USA}


\begin{abstract}

This letter presents the concept of Total Internal Reflection metasurface (TIR-MS) which supports the realization of structure-embedded subwavelength acoustic shields for elastic waves propagating in thin waveguides. The proposed metasurface design exploits extreme phase gradients, implemented via locally resonant elements, in order to achieve operating conditions that are largely beyond the critical angle. Such artificial discontinuity is capable of producing complete reflection of the incoming waves regardless of the specific angle of incidence. From a practical perspective, the TIR-MS behaves as a sound hard barrier that is impenetrable to long-wavelength modes at a selected frequency. The TIR metasurface concept is first conceived for a flat interface embedded in a rectangular waveguide and designed to block longitudinal $S_0$-type guided modes. Then, it is extended to circular plates in order to show how enclosed areas can be effectively shielded by incoming waves. Given the same underlying physics, an equivalent dynamic behavior was also numerically and experimentally illustrated for flexural $A_0$-type guided modes. This study shows numerical and experimental evidence that, when the metasurface is excited at the target frequency, significant vibration isolation can be achieved in presence of waves having any arbitrary angle of incidence. These results open interesting paths to achieve vibration isolation and energy filtering in certain prototypical structures of interest for practical engineering applications.

\end{abstract}

\maketitle

\section{Introduction}

The concept of metasurface has recently emerged as a powerful approach to achieve compact and subwavelength devices for wave manipulation. The fundamental idea was first pioneered in optics \cite{Yu,Ni,Grady,Pfeiffer,Aieta,Kang,Sun,Huang} and later extended to acoustics \cite{Li1,Li2,Mei,Tang,Yuan,Zhao1,Zhao2,yifan,Tang2,Li5}. At the basis of the metasurface design lies the concept of Generalized Snell's Law (GSL) \cite{Yu,YuRev} that allows predicting the anomalous refraction across interfaces characterized by a phase gradient. Metasurfaces are rather versatile objects given the variety of phase-shift profiles that can be encoded in them and that lead to remarkable and unconventional wave manipulation effects. A few examples include bending wave fields (either light or sound) in arbitrary shapes \cite{Ni,Huang}, converting propagating into surface modes \cite{Sun}, and designing ultra-thin lenses \cite{Aieta,Kang}.

Only very recently, this concept was extended for application to elastodynamics \cite{ZhuPRL,EMS1,EMS2,EMS3,EMS4,EMS5,EMS6,EMS7} in order to control the refraction of elastic waves in solid waveguides. In their initial study, Zhu \textit{et al.} \cite{ZhuPRL} explored the behavior of an elastic metasurface in transmission mode and investigated the possibility of controlling the angle of refraction and the shape of an incoming planar wave front. 

In traditional problems of wave transmission across an interface between dissimilar materials, both the angles of reflection and refraction are controlled by either the angle of incidence of the incoming wave or by the impedance mismatch between the two materials. For a given material selection, it is well-known that upon increasing the angle of incidence of the incoming wave a condition will be reached in which the wave cannot propagate into the second material (i.e. no refracted wave solutions can exist). This threshold value of the incident angle is referred to as \textit{critical angle}. Typically, at the first critical angle the angle of refraction is $90^{\circ}$, therefore the wave travels along the interface with no transmission into the second material. If the angle of incidence is further increased beyond the critical value, the wave is entirely reflected into the same half space giving rise to a phenomenon known as \textit{total internal reflection} (TIR). In analogy with this classical behavior of wave propagation through an interface, we highlight that also a metasurface can achieve critical angle conditions. 

In this letter, we explore both theoretically and experimentally the design and performance of an elastic metasurface that is explicitly designed to operate in the TIR regime. Such design is particularly interesting because it provides a fully passive approach to achieve embedded subwavelength sound-hard barriers capable of blocking the propagation of elastic waves in the host waveguide. It is anticipated that this effect could be exploited in a variety of devices such as, for example, structural notch filters or, more in general, to achieve vibration isolation and control. 

In the following, we present the general conditions and a possible design necessary to achieve a TIR metasurface. We first validate numerically the design (on either a rectangular or a circular plate), then we present an experimental validation for the case of a circular plate assembled in a prototypical structure and illustrate how vibration isolation can be effectively achieved under highly subwavelength excitation conditions.

\section{Design of the metasurface for total internal reflection}
According to the Generalized Snell's Law (GSL) \cite{Yu,ZhuPRL}, when an interface between two (either identical or different) media is encoded with a phase gradient $d\phi / dy$, the direction of the refracted beam $\theta_t$ can be related to the incident angle $\theta_i$ as follows:
\begin{align}
\frac{sin(\theta_t)}{\lambda_t}- \frac{sin(\theta_i)}{\lambda_i}=\frac{1}{2\pi}\frac{d\phi}{dy}\label{eqn1}
\end{align}
Given that in order to achieve propagating conditions across the interface (i.e. real-valued wave numbers $k_t=2\pi/\lambda_t$) it must be $-1<sin(\theta_t)<1$, any choice of the phase gradient that violates this condition will result in total internal reflection. Substituting this condition into Eq. (\ref{eqn1}), we obtain that in order to achieve TIR the phase gradient must satisfy the inequality $\frac{d\phi}{dy} \ge \frac{4\pi}{\lambda}$. Additional requirements on the amplitude of the units' transfer function should also be satisfied as clarified in the supplementary material. 

Note that in locally-resonant metasurfaces, the phase shift gradient is discretized in piecewise constant segments, each corresponding to the fixed phase jump produced by a given unit cell. Therefore, the minimum number of required unit cells to represent a non-zero $d\phi/dy$ phase gradient is two. Under deep subwavelength conditions, this coarse discretization of the phase gradient can still yield satisfactory performance. In support of this statement, we note that the experimental validation presented below uses indeed only two units to discretize the phase gradient. Nevertheless, we mention that the level of discretization is application dependent and should be assessed depending especially on the operating wavelength conditions. As a general rule, a smoother discretization (that is more unit cells per $2\pi$ phase range) is needed as the wavelength decreases. When the discretization is too coarse, slit-like diffraction effects are generated between adjacent units and tend to deform the wave front.

\begin{figure}
	\includegraphics[scale=0.4]{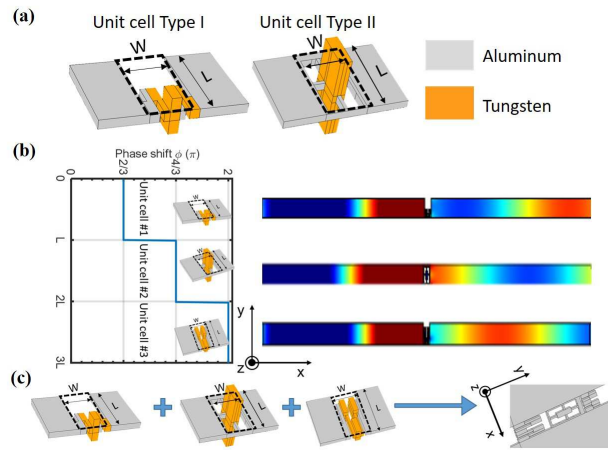}
	\caption{ Fundamental resonating unit cells of the TIR metasurface. (a) shows the schematics of the two unit cells used in our approach. The unit cells are designed based on either a space-coiling or a spring-mass approach and employ different materials. (b) shows the in-plane displacement component of the $S_0$ wave traveling from left to right. The transmitted field has comparable amplitude and the required phase shift (in increments of $120^{\circ}$). (c) The three selected unit cells can be assembled in a supercell that covers a $2\pi$ phase shift. Then, the supercell is periodically repeated to assemble the complete metasurface.} \label{Fig1}
\end{figure}

In this study, the metasurface employed a locally resonant unit (Fig.\ref{Fig1}a) made by a rectangular block ($W=40 \ mm \times L=80\ mm \times t=8\ mm$) connected to a more elaborate internal structure. Depending on the specific values of the design parameters, the unit provided a phase shift under extreme subwavelength conditions. It is worth noting that both add-on masses and dissimilar materials (in the numerical model chosen to be aluminum and tungsten) were employed in order to tune the resonance of the units while maintaining a compact design. The internal structure was designed either according to the idea of space-coiling or of a simple spring-mass system (see details in Supplementary Material). From an effective medium perspective, the resonating unit cells can be viewed as highly refractive index media (phase velocity $c \propto \sqrt{\frac{E_{eff}}{M_{eff}}}$). In order to cover the $2\pi$ phase range under extreme subwavelength conditions, a general approach is to reduce the effective stiffness and increase the effective mass. The effective stiffness can be controlled by tuning either the \textit{path length}, for the space-coiling design, or the cross-sectional dimensions of the connecting beam, for the spring-mass type. The effective mass can be controlled by either using high density materials or by adding discrete masses. The independent control of the effective stiffness and mass also facilitate the mechanical impedance compensation (impedance $\propto \sqrt{M_{eff}E_{eff}}$) that is needed to maintain the same transmission coefficient across each unit cell. More details on the design strategy of elastic locally-resonant unit cells for metausurface applications can be found in [\cite{ZhuPRL}].

A preliminary set of simulations was performed to evaluate the dynamic behavior of different unit cell designs. Phase and amplitude of the corresponding transfer functions were numerically evaluated at the target frequency for different geometric parameters. These results allowed the formation of a database of unit cells having different transfer functions (or, more properly, different amplitudes and phases) at the target frequency. Based on either the geometries available in this database or on the target phase gradient required to the metasurface, proper units were selected. 

We started our analysis assembling a metasurface capable of total internal reflection under a $S_0$ excitation at normal incidence and at a target frequency $f=4.1$ kHz. The first step of the design required assembling, from individual resonating units (Fig.\ref{Fig1}c), a supercell capable of providing the necessary phase gradient and amplitude profile. Three basic units having a transmission coefficient of $\approx 0.32$ and a phase shift difference of $2\pi/3$ between adjacent units were selected (see supplementary material). The effect on the phase shift provided by each unit on the transmitted $S_0$ mode is shown in Fig.\ref{Fig1}b. The plots show the selected geometry for the resonant unit and the corresponding in-plane particle displacement pattern $U_x(x,y)$ at the neutral plane. Results clearly show a cumulative phase shift covering the whole $2\pi$ range. The latter is a required condition to guarantee that, when multiple supercells are assembled in a periodic pattern to form the metasurface, the phase gradient evolves smoothly in increments that are integer multiple of $2\pi$.

\begin{figure}
	\includegraphics[scale=0.42]{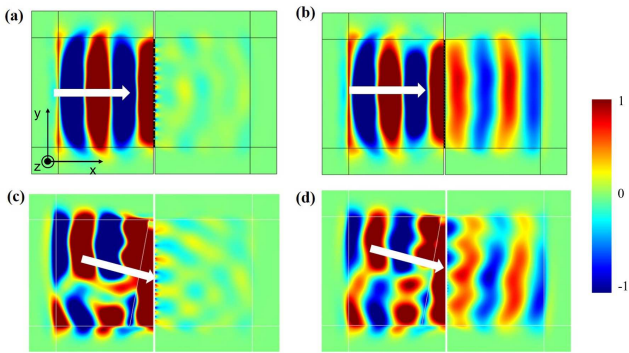}
	\caption{Numerical simulations showing the response of the TIR metasurface embedded in a thin rectangular waveguide. (a) and (c) show the in-plane particle displacement component produced by the $S_0$ wave field under a $S_0$ input wave at normal and oblique incidence, respectively. (b) and (d) shows results analogous to (a) and (c) when the TIR-MS is replaced by a metasurface without phase gradient.} \label{Fig2}
\end{figure}

\section{Numerical Simulations}

To illustrate the validity of the design and the performance of the metasurface in terms of total internal reflection, we embedded the TIR metasurface in a thin aluminum rectangular waveguide having thickness $t=8\ mm$ (Fig.\ref{Fig1}c). The corresponding numerical model was assembled and solved in a commercial finite element software (COMSOL Multiphysics\texttrademark). The target frequency was selected at $f=4.1$ kHz and the phase increment between two adjacent units was $\det\phi=2\pi/3$. This condition fulfilled the requirement necessary to achieve TIR effects and provided a very subwavelength design as demonstrated by the ratio $\lambda /\Delta x\approx 12.5$, where $\Delta x$ is the width of the metasurface. The metasurface was excited by a $S_0$ planar wave traveling from left to right (i.e. positive x-direction) under two conditions: 1) normal incidence, and 2) oblique incidence ($\theta_i=30^{\circ}$). 

Figure~\ref{Fig2}a and c show the in-plane displacement component produced by the $S_0$ guided mode after interaction with the TIR-MS at normal and oblique incidence (i.e. $\theta_i=30^{\circ}$), respectively. The superimposed white arrow indicates the incident wave direction. Results clearly show that most of the incident energy is back reflected while just a minor portion is refracted through the metasurface. This transmitted scattered field is due to the discretized nature of the interface (i.e. the openings between adjacent units) and resemble a typical slit diffraction effect.

To further illustrate the performance of the TIR metasurface, we simulated the propagation conditions in an equivalent waveguide (indicated below as reference configuration) where the TIR-MS was replaced by a geometrically equivalent metasurface. The latter interface had the same transmission amplitude than the TIR-MS but it did not provide any phase shift gradient. Such metasurface was assembled by repeating only one of the basic unit cells along the $y-$direction. The intent of this comparison is to show that the back reflection of the incident wave is not due to a simple impedance mismatch but instead to the actual design of the TIR interface.
The numerical results of the reference configuration are shown in Fig.\ref{Fig2}b and d for the two incidence angles. The results are still presented in terms of in-plane particle displacement of the $S_0$ wave field. By direct comparison with Fig.\ref{Fig2}a and c, it is evident that although some back reflections due to the impedance mismatch certainly occurs, only the TIR design is capable of blocking the wave under this very challenging subwavelength condition.

\begin{figure}
	\includegraphics[scale=0.42]{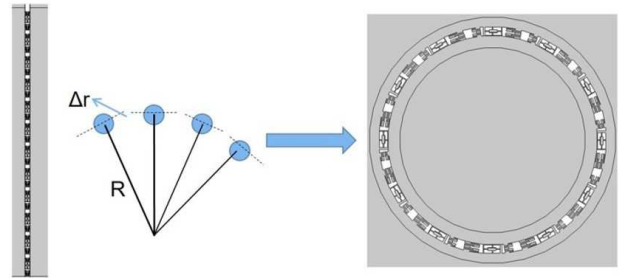}
	\caption{(a) Conceptual schematic illustrating how the flat metasurface design can be adapted to create enclosed areas effectively shielded from unwanted incoming mechanical energy. } \label{Fig3}
\end{figure}
From the results presented above, it appears that the TIR-MS is particularly well suited for blocking the propagation of wave fronts under highly subwavelength conditions and, eventually, to create analog notch filters that block a selected frequency. To illustrate this specific capability of the TIR-MS, we show the application of this concept to a thin circular waveguide. The underlying idea being that one wants to create fully enclosed areas shielded by incoming waves at a given frequency but with an arbitrary angle of incidence.
In order to directly extend the results from the previous section, we can envision bending the flat TIR-MS into a circular pattern closed on itself as shown in Fig.\ref{Fig3}. For completeness, the numerical results for this configuration are summarized in the supplementary materials. It is also important to note that, although the design of the TIR-MS was illustrated above with respect to $S_0$ excitation, equivalent design strategy and operating principles are applicable to $A_0$ excitation. An example of circular TIR-MS designed to block $A_0$ modes is provided in the supplementary material.

\begin{figure}
	\includegraphics[scale=0.42]{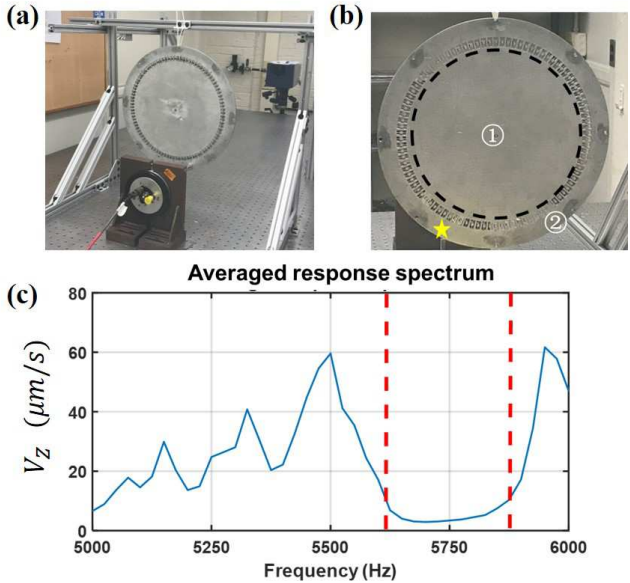}
	\caption{Experimental setup consisting of a circular plate with an embedded circular TIR-MS. The plate was tested under free boundary conditions and the excitation was provided by a piezoelectric shaker connected to the outer region, as indicated in (a). (a) and (b) show the rear and front view of the set-up. (c) shows the spatially-averaged (out-of-plane) velocity spectrum obtained from all the points internal to the metasurface (i.e. zone \textcircled{1}). The spectrum was measured under white noise excitation with a 5-6 kHz bandwidth. The measurements show a strong reduction in the transmitted energy within zone \textcircled{1} around $f=5.750$ kHz, which is the approximate operating frequency of the metasurface.} \label{Fig4}
\end{figure}

\section{Experimental Validation}

In order to validate the TIR-MS concept and evaluate its performance, we built an experimental test bed and measured its response using laser vibrometry. In particular, we reproduced the specific case of a circular TIR-MS designed to isolate its most inner area from an incoming $A_0$ mode. To facilitate the fabrication and testing, we re-designed the metasurface by rescaling the basic unit, and by using aluminum and steel as materials. The rescaling procedure resulted in a target frequency of $5.5$ kHz for the $A_0$ mode (the details of the geometry are reported in supplementary material). The circular TIR-MS was fabricated by waterjet cutting the units starting from a uniform flat plate and then gluing the add-on masses. In this design, the supercell consisted of only two unit cells, each producing a $\pi$ phase shift. As previously mentioned, this is the minimum number of unit cells that can be used to discretize the phase gradient and represents the most challenging condition in terms of diffraction effects.
\begin{figure}
	\includegraphics[scale=0.43]{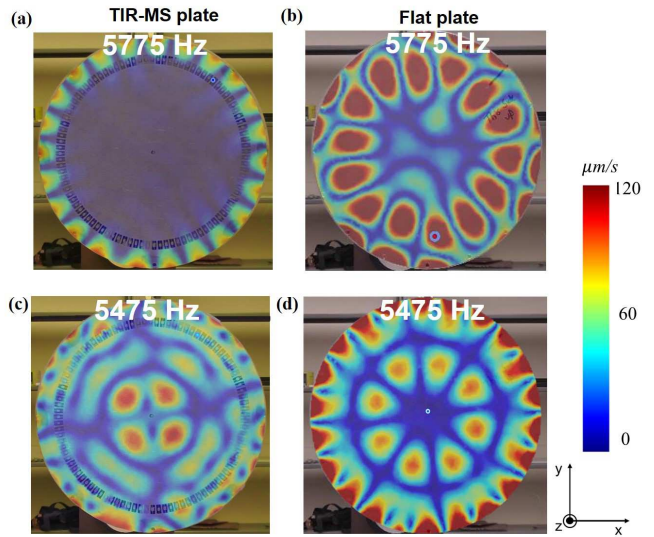}
	\caption{Experimental results. (a) shows the response of the TIR-MS plate at the target frequency $f=5.775$ kHz inside the spectrum dip (see Fig.\ref{Fig4}c). The inner region \textcircled{1} is almost completely isolated from incoming waves. For comparison, (b) shows the response of a flat plate under the same excitation condition. It is clear that in this case the incoming waves can travel through the inner region and establish traditional standing wave patterns. (c) shows the response of the MS-plate at a detuned frequency $f=5.475$ kHz outside the operating range, and (c) shows the response of the flat plate under the same excitation condition. } \label{Fig5}
\end{figure}

The test setup is shown in Fig.\ref{Fig4}. The rear and front views of the whole setup are provided in Fig.\ref{Fig4}a and b. The plate was suspended from an aluminum frame in order to provide free boundary conditions while a piezoelectric shaker was connected to the outer area (see $\star$ in Fig.\ref{Fig4}b) in order to provide the necessary dynamic excitation. The goal of the metasurface was to dynamically isolate the internal region (zone \textcircled{1}) from the incoming waves generated in zone \textcircled{2}. The response of the plate was measured under white noise excitation (bandwidth 5-6 kHz) by using a PSV-500 laser vibrometer. Figure~\ref{Fig4}c presents the amplitude of the spatially-averaged velocity spectrum of the internal region (zone \textcircled{1}). The large drop in the amplitude of the spectrum in the frequency band $5.65-5.85$ kHz identifies the region in which the metasurface is operating (around a center frequency of approximately 5.75 kHz). This frequency is only 250 Hz off the numerically predicted target frequency, that is a $4 \%$ error. This is considered a negligible discrepancy in this frequency range and it is associated with fabrication tolerances on both the individual unit cell slots and the add-on masses.

Figure \ref{Fig5} summarizes the results of the experimental measurements. In particular, Fig. \ref{Fig5}a and b show the response of the TIR-MS and of the reference flat plates at the frequency $f=5.775$ kHz within the operating range of the metasurface (see dashed lines in Fig.\ref{Fig4}c). The response is measure in terms of out-of-plane velocity $V_{z}$ amplitude distribution. As expected, in the TIR-MS plate the vibrational energy is confined outside the metasurface, resulting in the formation of a lobbed pattern while leaving the internal region almost completely unaffected. On the contrary, in the case of the flat plate (Fig.\ref{Fig5}b) the excitation produces a pattern resembling a global vibration mode for the most part localized in the region equivalent to the internal area of the metasurface. In order to provide a quantitative comparison of the effect of the metasurface in terms of energy isolation, we calculated the amplitude of the spatially-averaged velocity spectrum in zone \textcircled{1} and \textcircled{2}, labeled as $V_{z1}$ and $V_{z2}$. For the TIR-MS plate $V^{MS}_{z1}=5.29 \ \mu m/s$ and $V^{MS}_{z2}=52.84 \ \mu m/s$, while for the flat plate plate $V^{flat}_{z1}=182.11 \ \mu m/s$ and $V^{flat}_{z2}=192.14 \ \mu m/s$. The direct comparison of this quantities shows that most of the energy is confined in zone \textcircled{2} in the case of the TIR-MS while it reaches zone \textcircled{1} for the flat plate.
	
For completeness, we also report the response of the two plates for a slightly detuned frequency $f=5.475$ kHz, that is outside the operating range of the TIR-MS (i.e. outside the dashed lines in Fig.\ref{Fig4}c). A direct comparison of the velocity spectra (Fig.\ref{Fig5}c and d) shows that the vibrational energy can reach zone \textcircled{1} in both plates given that the excitation frequency cannot activate the metasurface. The resulting vibration pattern shows differences in the two plates due to the local discontinuity produced by the MS, however in both cases the response has the classical hallmarks of a global (standing wave) mode.

\section{Conclusions}
We have presented and experimentally validated the concept of total internal reflection elastic metasurface (TIR-MS) which is capable of highly subwavelength dynamic isolation. The TIR-MS relies on the existence of a critical phase shift gradient which guarantees that incident wave fronts at any arbitrary angle cannot be transmitted across the engineered interface. The specific metasurface design employed locally resonant units capable of achieving large phase shifts covering the whole $2\pi$ range under highly subwavelength conditions. Effective wave control was shown first numerically and then experimentally on different types of thin elastic waveguides. While traditional metasurfaces are typically designed to be transparent and to refract waves in a controlled manner, the TIR-MS is designed to be opaque or, ideally, fully reflective. Results show that TIR-MS is particularly well suited for blocking the propagation of wave fronts with an arbitrary angle of incidence and at a selected frequency. This aspect of the TIR-MS is particularly attractive for vibration isolation applications where the objective is to block the propagation of low frequency (i.e. long wavelength) waves while preserving the structural role of the host system. Given the narrow band characteristic of the resonant metasurface design, it is expected that the TIR-MS could also be employed as an analog notch filter to block a selected frequency. 

\section{Supplementary Material}

See Supplementary Material for discussions on the design strategy, the geometric parameters for selected unit cells, and additional numerical results of the circular TIR metasurface.

\section{Acknowledgments}
The authors gratefully acknowledge the financial support of the Sandia National Laboratory under the Academic Alliance program, grant \#1847039.


\begin{thebibliography}{27}%
	\makeatletter
	\providecommand \@ifxundefined [1]{%
		\@ifx{#1\undefined}
	}%
	\providecommand \@ifnum [1]{%
		\ifnum #1\expandafter \@firstoftwo
		\else \expandafter \@secondoftwo
		\fi
	}%
	\providecommand \@ifx [1]{%
		\ifx #1\expandafter \@firstoftwo
		\else \expandafter \@secondoftwo
		\fi
	}%
	\providecommand \natexlab [1]{#1}%
	\providecommand \enquote  [1]{``#1''}%
	\providecommand \bibnamefont  [1]{#1}%
	\providecommand \bibfnamefont [1]{#1}%
	\providecommand \citenamefont [1]{#1}%
	\providecommand \href@noop [0]{\@secondoftwo}%
	\providecommand \href [0]{\begingroup \@sanitize@url \@href}%
	\providecommand \@href[1]{\@@startlink{#1}\@@href}%
	\providecommand \@@href[1]{\endgroup#1\@@endlink}%
	\providecommand \@sanitize@url [0]{\catcode `\\12\catcode `\$12\catcode
		`\&12\catcode `\#12\catcode `\^12\catcode `\_12\catcode `\%12\relax}%
	\providecommand \@@startlink[1]{}%
	\providecommand \@@endlink[0]{}%
	\providecommand \url  [0]{\begingroup\@sanitize@url \@url }%
	\providecommand \@url [1]{\endgroup\@href {#1}{\urlprefix }}%
	\providecommand \urlprefix  [0]{URL }%
	\providecommand \Eprint [0]{\href }%
	\providecommand \doibase [0]{http://dx.doi.org/}%
	\providecommand \selectlanguage [0]{\@gobble}%
	\providecommand \bibinfo  [0]{\@secondoftwo}%
	\providecommand \bibfield  [0]{\@secondoftwo}%
	\providecommand \translation [1]{[#1]}%
	\providecommand \BibitemOpen [0]{}%
	\providecommand \bibitemStop [0]{}%
	\providecommand \bibitemNoStop [0]{.\EOS\space}%
	\providecommand \EOS [0]{\spacefactor3000\relax}%
	\providecommand \BibitemShut  [1]{\csname bibitem#1\endcsname}%
	\let\auto@bib@innerbib\@empty
	\bibitem [{\citenamefont {Yu}\ \emph {et~al.}(2011)\citenamefont {Yu},
		\citenamefont {Genevet}, \citenamefont {Kats}, \citenamefont {Aieta},
		\citenamefont {Tetienne}, \citenamefont {Capasso},\ and\ \citenamefont
		{Gaburro}}]{Yu}%
	\BibitemOpen
	\bibfield  {author} {\bibinfo {author} {\bibfnamefont {N.}~\bibnamefont
			{Yu}}, \bibinfo {author} {\bibfnamefont {P.}~\bibnamefont {Genevet}},
		\bibinfo {author} {\bibfnamefont {M.~A.}\ \bibnamefont {Kats}}, \bibinfo
		{author} {\bibfnamefont {F.}~\bibnamefont {Aieta}}, \bibinfo {author}
		{\bibfnamefont {J.-P.}\ \bibnamefont {Tetienne}}, \bibinfo {author}
		{\bibfnamefont {F.}~\bibnamefont {Capasso}}, \ and\ \bibinfo {author}
		{\bibfnamefont {Z.}~\bibnamefont {Gaburro}},\ }\href@noop {} {\bibfield
		{journal} {\bibinfo  {journal} {Science}\ }\textbf {\bibinfo {volume}
			{334}},\ \bibinfo {pages} {333} (\bibinfo {year} {2011})}\BibitemShut
	{NoStop}%
	\bibitem [{\citenamefont {Ni}\ \emph {et~al.}(2012)\citenamefont {Ni},
		\citenamefont {Emani}, \citenamefont {Kildishev}, \citenamefont
		{Boltasseva},\ and\ \citenamefont {Shalaev}}]{Ni}%
	\BibitemOpen
	\bibfield  {author} {\bibinfo {author} {\bibfnamefont {X.}~\bibnamefont
			{Ni}}, \bibinfo {author} {\bibfnamefont {N.~K.}\ \bibnamefont {Emani}},
		\bibinfo {author} {\bibfnamefont {A.~V.}\ \bibnamefont {Kildishev}}, \bibinfo
		{author} {\bibfnamefont {A.}~\bibnamefont {Boltasseva}}, \ and\ \bibinfo
		{author} {\bibfnamefont {V.~M.}\ \bibnamefont {Shalaev}},\ }\href@noop {}
	{\bibfield  {journal} {\bibinfo  {journal} {Science}\ }\textbf {\bibinfo
			{volume} {335}},\ \bibinfo {pages} {427} (\bibinfo {year}
		{2012})}\BibitemShut {NoStop}%
	\bibitem [{\citenamefont {Grady}\ \emph {et~al.}(2013)\citenamefont {Grady},
		\citenamefont {Heyes}, \citenamefont {Chowdhury}, \citenamefont {Zeng},
		\citenamefont {Reiten}, \citenamefont {Azad}, \citenamefont {Taylor},
		\citenamefont {Dalvit},\ and\ \citenamefont {Chen}}]{Grady}%
	\BibitemOpen
	\bibfield  {author} {\bibinfo {author} {\bibfnamefont {N.~K.}\ \bibnamefont
			{Grady}}, \bibinfo {author} {\bibfnamefont {J.~E.}\ \bibnamefont {Heyes}},
		\bibinfo {author} {\bibfnamefont {D.~R.}\ \bibnamefont {Chowdhury}}, \bibinfo
		{author} {\bibfnamefont {Y.}~\bibnamefont {Zeng}}, \bibinfo {author}
		{\bibfnamefont {M.~T.}\ \bibnamefont {Reiten}}, \bibinfo {author}
		{\bibfnamefont {A.~K.}\ \bibnamefont {Azad}}, \bibinfo {author}
		{\bibfnamefont {A.~J.}\ \bibnamefont {Taylor}}, \bibinfo {author}
		{\bibfnamefont {D.~A.~R.}\ \bibnamefont {Dalvit}}, \ and\ \bibinfo {author}
		{\bibfnamefont {H.-T.}\ \bibnamefont {Chen}},\ }\href@noop {} {\bibfield
		{journal} {\bibinfo  {journal} {Science}\ }\textbf {\bibinfo {volume}
			{340}},\ \bibinfo {pages} {1304} (\bibinfo {year} {2013})}\BibitemShut
	{NoStop}%
	\bibitem [{\citenamefont {Pfeiffer}\ and\ \citenamefont
		{Grbic}(2013)}]{Pfeiffer}%
	\BibitemOpen
	\bibfield  {author} {\bibinfo {author} {\bibfnamefont {C.}~\bibnamefont
			{Pfeiffer}}\ and\ \bibinfo {author} {\bibfnamefont {A.}~\bibnamefont
			{Grbic}},\ }\href@noop {} {\bibfield  {journal} {\bibinfo  {journal} {Phys.
				Rev. Lett.}\ }\textbf {\bibinfo {volume} {110}},\ \bibinfo {pages} {197401}
		(\bibinfo {year} {2013})}\BibitemShut {NoStop}%
	\bibitem [{\citenamefont {Aieta}\ \emph {et~al.}(2012)\citenamefont {Aieta},
		\citenamefont {Genevet}, \citenamefont {Kats}, \citenamefont {Yu},
		\citenamefont {Blanchard}, \citenamefont {Gaburro},\ and\ \citenamefont
		{Capasso}}]{Aieta}%
	\BibitemOpen
	\bibfield  {author} {\bibinfo {author} {\bibfnamefont {F.}~\bibnamefont
			{Aieta}}, \bibinfo {author} {\bibfnamefont {P.}~\bibnamefont {Genevet}},
		\bibinfo {author} {\bibfnamefont {M.~A.}\ \bibnamefont {Kats}}, \bibinfo
		{author} {\bibfnamefont {N.}~\bibnamefont {Yu}}, \bibinfo {author}
		{\bibfnamefont {R.}~\bibnamefont {Blanchard}}, \bibinfo {author}
		{\bibfnamefont {Z.}~\bibnamefont {Gaburro}}, \ and\ \bibinfo {author}
		{\bibfnamefont {F.}~\bibnamefont {Capasso}},\ }\href@noop {} {\bibfield
		{journal} {\bibinfo  {journal} {Nano. Lett.}\ }\textbf {\bibinfo {volume}
			{12}},\ \bibinfo {pages} {4932} (\bibinfo {year} {2012})}\BibitemShut
	{NoStop}%
	\bibitem [{\citenamefont {Kang}\ \emph {et~al.}(2012)\citenamefont {Kang},
		\citenamefont {Feng}, \citenamefont {Wang},\ and\ \citenamefont {Li}}]{Kang}%
	\BibitemOpen
	\bibfield  {author} {\bibinfo {author} {\bibfnamefont {M.}~\bibnamefont
			{Kang}}, \bibinfo {author} {\bibfnamefont {T.}~\bibnamefont {Feng}}, \bibinfo
		{author} {\bibfnamefont {H.-T.}\ \bibnamefont {Wang}}, \ and\ \bibinfo
		{author} {\bibfnamefont {J.}~\bibnamefont {Li}},\ }\href@noop {} {\bibfield
		{journal} {\bibinfo  {journal} {Opt. Express}\ }\textbf {\bibinfo {volume}
			{20}},\ \bibinfo {pages} {15882} (\bibinfo {year} {2012})}\BibitemShut
	{NoStop}%
	\bibitem [{\citenamefont {Sun}\ \emph {et~al.}(2012)\citenamefont {Sun},
		\citenamefont {He}, \citenamefont {Xiao}, \citenamefont {Xu}, \citenamefont
		{Li},\ and\ \citenamefont {Zhou}}]{Sun}%
	\BibitemOpen
	\bibfield  {author} {\bibinfo {author} {\bibfnamefont {S.}~\bibnamefont
			{Sun}}, \bibinfo {author} {\bibfnamefont {Q.}~\bibnamefont {He}}, \bibinfo
		{author} {\bibfnamefont {S.}~\bibnamefont {Xiao}}, \bibinfo {author}
		{\bibfnamefont {Q.}~\bibnamefont {Xu}}, \bibinfo {author} {\bibfnamefont
			{X.}~\bibnamefont {Li}}, \ and\ \bibinfo {author} {\bibfnamefont
			{L.}~\bibnamefont {Zhou}},\ }\href@noop {} {\bibfield  {journal} {\bibinfo
			{journal} {Nat. Mater.}\ }\textbf {\bibinfo {volume} {11}},\ \bibinfo {pages}
		{426 } (\bibinfo {year} {2012})}\BibitemShut {NoStop}%
	\bibitem [{\citenamefont {Huang}\ \emph {et~al.}(2012)\citenamefont {Huang},
		\citenamefont {Chen}, \citenamefont {Muhlenbernd}, \citenamefont {Li},
		\citenamefont {Bai}, \citenamefont {Tan}, \citenamefont {Jin}, \citenamefont
		{Zentgraf},\ and\ \citenamefont {Zhang}}]{Huang}%
	\BibitemOpen
	\bibfield  {author} {\bibinfo {author} {\bibfnamefont {L.}~\bibnamefont
			{Huang}}, \bibinfo {author} {\bibfnamefont {X.}~\bibnamefont {Chen}},
		\bibinfo {author} {\bibfnamefont {H.}~\bibnamefont {Muhlenbernd}}, \bibinfo
		{author} {\bibfnamefont {G.}~\bibnamefont {Li}}, \bibinfo {author}
		{\bibfnamefont {B.}~\bibnamefont {Bai}}, \bibinfo {author} {\bibfnamefont
			{Q.}~\bibnamefont {Tan}}, \bibinfo {author} {\bibfnamefont {G.}~\bibnamefont
			{Jin}}, \bibinfo {author} {\bibfnamefont {T.}~\bibnamefont {Zentgraf}}, \
		and\ \bibinfo {author} {\bibfnamefont {S.}~\bibnamefont {Zhang}},\
	}\href@noop {} {\bibfield  {journal} {\bibinfo  {journal} {Nano. Lett.}\
		}\textbf {\bibinfo {volume} {12}},\ \bibinfo {pages} {5750 } (\bibinfo {year}
		{2012})}\BibitemShut {NoStop}%
	\bibitem [{\citenamefont {Li}\ \emph {et~al.}(2013)\citenamefont {Li},
		\citenamefont {Liang}, \citenamefont {Gu}, \citenamefont {Zou},\ and\
		\citenamefont {Cheng}}]{Li1}%
	\BibitemOpen
	\bibfield  {author} {\bibinfo {author} {\bibfnamefont {Y.}~\bibnamefont
			{Li}}, \bibinfo {author} {\bibfnamefont {B.}~\bibnamefont {Liang}}, \bibinfo
		{author} {\bibfnamefont {Z.}~\bibnamefont {Gu}}, \bibinfo {author}
		{\bibfnamefont {X.}~\bibnamefont {Zou}}, \ and\ \bibinfo {author}
		{\bibfnamefont {J.}~\bibnamefont {Cheng}},\ }\href@noop {} {\bibfield
		{journal} {\bibinfo  {journal} {Sci. Rep.}\ }\textbf {\bibinfo {volume} {3}}
		(\bibinfo {year} {2013})}\BibitemShut {NoStop}%
	\bibitem [{\citenamefont {Li}\ \emph {et~al.}(2014)\citenamefont {Li},
		\citenamefont {Jiang}, \citenamefont {Li}, \citenamefont {Liang},
		\citenamefont {Zou}, \citenamefont {Yin},\ and\ \citenamefont {Cheng}}]{Li2}%
	\BibitemOpen
	\bibfield  {author} {\bibinfo {author} {\bibfnamefont {Y.}~\bibnamefont
			{Li}}, \bibinfo {author} {\bibfnamefont {X.}~\bibnamefont {Jiang}}, \bibinfo
		{author} {\bibfnamefont {R.-q.}\ \bibnamefont {Li}}, \bibinfo {author}
		{\bibfnamefont {B.}~\bibnamefont {Liang}}, \bibinfo {author} {\bibfnamefont
			{X.-y.}\ \bibnamefont {Zou}}, \bibinfo {author} {\bibfnamefont {L.-l.}\
			\bibnamefont {Yin}}, \ and\ \bibinfo {author} {\bibfnamefont {J.-c.}\
			\bibnamefont {Cheng}},\ }\href@noop {} {\bibfield  {journal} {\bibinfo
			{journal} {Phys. Rev. Applied}\ }\textbf {\bibinfo {volume} {2}},\ \bibinfo
		{pages} {064002} (\bibinfo {year} {2014})}\BibitemShut {NoStop}%
	\bibitem [{\citenamefont {Mei}\ and\ \citenamefont {Wu}(2014)}]{Mei}%
	\BibitemOpen
	\bibfield  {author} {\bibinfo {author} {\bibfnamefont {J.}~\bibnamefont
			{Mei}}\ and\ \bibinfo {author} {\bibfnamefont {Y.}~\bibnamefont {Wu}},\
	}\href@noop {} {\bibfield  {journal} {\bibinfo  {journal} {New. J. Phys.}\
		}\textbf {\bibinfo {volume} {16}},\ \bibinfo {pages} {123007} (\bibinfo
		{year} {2014})}\BibitemShut {NoStop}%
	\bibitem [{\citenamefont {Tang}\ \emph {et~al.}(2014)\citenamefont {Tang},
		\citenamefont {Qiu}, \citenamefont {Ke}, \citenamefont {Lu}, \citenamefont
		{Ye},\ and\ \citenamefont {Liu}}]{Tang}%
	\BibitemOpen
	\bibfield  {author} {\bibinfo {author} {\bibfnamefont {K.}~\bibnamefont
			{Tang}}, \bibinfo {author} {\bibfnamefont {C.}~\bibnamefont {Qiu}}, \bibinfo
		{author} {\bibfnamefont {M.}~\bibnamefont {Ke}}, \bibinfo {author}
		{\bibfnamefont {J.}~\bibnamefont {Lu}}, \bibinfo {author} {\bibfnamefont
			{Y.}~\bibnamefont {Ye}}, \ and\ \bibinfo {author} {\bibfnamefont
			{Z.}~\bibnamefont {Liu}},\ }\href@noop {} {\bibfield  {journal} {\bibinfo
			{journal} {Sci. Rep.}\ }\textbf {\bibinfo {volume} {4}} (\bibinfo {year}
		{2014})}\BibitemShut {NoStop}%
	\bibitem [{\citenamefont {Yuan}, \citenamefont {Cheng},\ and\ \citenamefont
		{Liu}(2015)}]{Yuan}%
	\BibitemOpen
	\bibfield  {author} {\bibinfo {author} {\bibfnamefont {B.}~\bibnamefont
			{Yuan}}, \bibinfo {author} {\bibfnamefont {Y.}~\bibnamefont {Cheng}}, \ and\
		\bibinfo {author} {\bibfnamefont {X.}~\bibnamefont {Liu}},\ }\href@noop {}
	{\bibfield  {journal} {\bibinfo  {journal} {Appl. Phys. Express.}\ }\textbf
		{\bibinfo {volume} {8}} (\bibinfo {year} {2015})}\BibitemShut {NoStop}%
	\bibitem [{\citenamefont {Zhao}\ \emph
		{et~al.}(2013{\natexlab{a}})\citenamefont {Zhao}, \citenamefont {Li},
		\citenamefont {Chen},\ and\ \citenamefont {Qiu}}]{Zhao1}%
	\BibitemOpen
	\bibfield  {author} {\bibinfo {author} {\bibfnamefont {J.}~\bibnamefont
			{Zhao}}, \bibinfo {author} {\bibfnamefont {B.}~\bibnamefont {Li}}, \bibinfo
		{author} {\bibfnamefont {Z.}~\bibnamefont {Chen}}, \ and\ \bibinfo {author}
		{\bibfnamefont {C.~W.}\ \bibnamefont {Qiu}},\ }\href@noop {} {\bibfield
		{journal} {\bibinfo  {journal} {Sci. Rep.}\ }\textbf {\bibinfo {volume} {3}}
		(\bibinfo {year} {2013}{\natexlab{a}})}\BibitemShut {NoStop}%
	\bibitem [{\citenamefont {Zhao}\ \emph
		{et~al.}(2013{\natexlab{b}})\citenamefont {Zhao}, \citenamefont {Li},
		\citenamefont {Chen},\ and\ \citenamefont {Qiu}}]{Zhao2}%
	\BibitemOpen
	\bibfield  {author} {\bibinfo {author} {\bibfnamefont {J.}~\bibnamefont
			{Zhao}}, \bibinfo {author} {\bibfnamefont {B.}~\bibnamefont {Li}}, \bibinfo
		{author} {\bibfnamefont {Z.~N.}\ \bibnamefont {Chen}}, \ and\ \bibinfo
		{author} {\bibfnamefont {C.-W.}\ \bibnamefont {Qiu}},\ }\href@noop {}
	{\bibfield  {journal} {\bibinfo  {journal} {Appl. Phys. Lett.}\ }\textbf
		{\bibinfo {volume} {103}} (\bibinfo {year} {2013}{\natexlab{b}})}\BibitemShut
	{NoStop}%
	\bibitem [{\citenamefont {Zhu}\ \emph {et~al.}(2015)\citenamefont {Zhu},
		\citenamefont {Zou}, \citenamefont {Li}, \citenamefont {Jiang}, \citenamefont
		{Tu}, \citenamefont {Liang},\ and\ \citenamefont {Cheng}}]{yifan}%
	\BibitemOpen
	\bibfield  {author} {\bibinfo {author} {\bibfnamefont {Y.}~\bibnamefont
			{Zhu}}, \bibinfo {author} {\bibfnamefont {X.}~\bibnamefont {Zou}}, \bibinfo
		{author} {\bibfnamefont {R.}~\bibnamefont {Li}}, \bibinfo {author}
		{\bibfnamefont {X.}~\bibnamefont {Jiang}}, \bibinfo {author} {\bibfnamefont
			{J.}~\bibnamefont {Tu}}, \bibinfo {author} {\bibfnamefont {B.}~\bibnamefont
			{Liang}}, \ and\ \bibinfo {author} {\bibfnamefont {J.-C.}\ \bibnamefont
			{Cheng}},\ }\href@noop {} {\bibfield  {journal} {\bibinfo  {journal} {Sci.
				Rep.}\ }\textbf {\bibinfo {volume} {5}} (\bibinfo {year} {2015})}\BibitemShut
	{NoStop}%
	\bibitem [{\citenamefont {Tang}\ \emph {et~al.}(2015)\citenamefont {Tang},
		\citenamefont {Qiu}, \citenamefont {Lu}, \citenamefont {Ke},\ and\
		\citenamefont {Liu}}]{Tang2}%
	\BibitemOpen
	\bibfield  {author} {\bibinfo {author} {\bibfnamefont {K.}~\bibnamefont
			{Tang}}, \bibinfo {author} {\bibfnamefont {C.}~\bibnamefont {Qiu}}, \bibinfo
		{author} {\bibfnamefont {J.}~\bibnamefont {Lu}}, \bibinfo {author}
		{\bibfnamefont {M.}~\bibnamefont {Ke}}, \ and\ \bibinfo {author}
		{\bibfnamefont {Z.}~\bibnamefont {Liu}},\ }\href@noop {} {\bibfield
		{journal} {\bibinfo  {journal} {J. Appl. Phys.}\ }\textbf {\bibinfo {volume}
			{117}} (\bibinfo {year} {2015})}\BibitemShut {NoStop}%
	\bibitem [{\citenamefont {Li}\ \emph {et~al.}(2015)\citenamefont {Li},
		\citenamefont {Jiang}, \citenamefont {Liang}, \citenamefont {Cheng},\ and\
		\citenamefont {Zhang}}]{Li5}%
	\BibitemOpen
	\bibfield  {author} {\bibinfo {author} {\bibfnamefont {Y.}~\bibnamefont
			{Li}}, \bibinfo {author} {\bibfnamefont {X.}~\bibnamefont {Jiang}}, \bibinfo
		{author} {\bibfnamefont {B.}~\bibnamefont {Liang}}, \bibinfo {author}
		{\bibfnamefont {J.-c.}\ \bibnamefont {Cheng}}, \ and\ \bibinfo {author}
		{\bibfnamefont {L.}~\bibnamefont {Zhang}},\ }\href@noop {} {\bibfield
		{journal} {\bibinfo  {journal} {Phys. Rev. Applied}\ }\textbf {\bibinfo
			{volume} {4}},\ \bibinfo {pages} {024003} (\bibinfo {year}
		{2015})}\BibitemShut {NoStop}%
	\bibitem [{\citenamefont {Yu}\ and\ \citenamefont {Capasso}(2014)}]{YuRev}%
	\BibitemOpen
	\bibfield  {author} {\bibinfo {author} {\bibfnamefont {N.}~\bibnamefont
			{Yu}}\ and\ \bibinfo {author} {\bibfnamefont {F.}~\bibnamefont {Capasso}},\
	}\href@noop {} {\bibfield  {journal} {\bibinfo  {journal} {Nat. Mater.}\
		}\textbf {\bibinfo {volume} {13}},\ \bibinfo {pages} {139 } (\bibinfo {year}
		{2014})}\BibitemShut {NoStop}%
	\bibitem [{\citenamefont {Zhu}\ and\ \citenamefont
		{Semperlotti}(2016)}]{ZhuPRL}%
	\BibitemOpen
	\bibfield  {author} {\bibinfo {author} {\bibfnamefont {H.}~\bibnamefont
			{Zhu}}\ and\ \bibinfo {author} {\bibfnamefont {F.}~\bibnamefont
			{Semperlotti}},\ }\href@noop {} {\bibfield  {journal} {\bibinfo  {journal}
			{Phys. Rev. Lett.}\ }\textbf {\bibinfo {volume} {117}},\ \bibinfo {pages}
		{034302} (\bibinfo {year} {2016})}\BibitemShut {NoStop}%
	\bibitem [{\citenamefont {Liu}\ \emph {et~al.}(2017)\citenamefont {Liu},
		\citenamefont {Liang}, \citenamefont {Liu}, \citenamefont {Diba},
		\citenamefont {Lamb},\ and\ \citenamefont {Li}}]{EMS1}%
	\BibitemOpen
	\bibfield  {author} {\bibinfo {author} {\bibfnamefont {Y.}~\bibnamefont
			{Liu}}, \bibinfo {author} {\bibfnamefont {Z.}~\bibnamefont {Liang}}, \bibinfo
		{author} {\bibfnamefont {F.}~\bibnamefont {Liu}}, \bibinfo {author}
		{\bibfnamefont {O.}~\bibnamefont {Diba}}, \bibinfo {author} {\bibfnamefont
			{A.}~\bibnamefont {Lamb}}, \ and\ \bibinfo {author} {\bibfnamefont
			{J.}~\bibnamefont {Li}},\ }\href@noop {} {\bibfield  {journal} {\bibinfo
			{journal} {Phys. Rev. Lett.}\ }\textbf {\bibinfo {volume} {119}},\ \bibinfo
		{pages} {034301} (\bibinfo {year} {2017})}\BibitemShut {NoStop}%
	\bibitem [{\citenamefont {Li}, \citenamefont {Xu},\ and\ \citenamefont
		{Tang}(2018)}]{EMS2}%
	\BibitemOpen
	\bibfield  {author} {\bibinfo {author} {\bibfnamefont {S.}~\bibnamefont
			{Li}}, \bibinfo {author} {\bibfnamefont {J.}~\bibnamefont {Xu}}, \ and\
		\bibinfo {author} {\bibfnamefont {J.}~\bibnamefont {Tang}},\ }\href@noop {}
	{\bibfield  {journal} {\bibinfo  {journal} {Applied Physics Letters}\
		}\textbf {\bibinfo {volume} {112}},\ \bibinfo {pages} {021903} (\bibinfo
		{year} {2018})}\BibitemShut {NoStop}%
	\bibitem [{\citenamefont {Su}, \citenamefont {Lu},\ and\ \citenamefont
		{Norris}(2018)}]{EMS3}%
	\BibitemOpen
	\bibfield  {author} {\bibinfo {author} {\bibfnamefont {X.}~\bibnamefont
			{Su}}, \bibinfo {author} {\bibfnamefont {Z.}~\bibnamefont {Lu}}, \ and\
		\bibinfo {author} {\bibfnamefont {A.~N.}\ \bibnamefont {Norris}},\
	}\href@noop {} {\bibfield  {journal} {\bibinfo  {journal} {Journal of Applied
				Physics}\ }\textbf {\bibinfo {volume} {123}},\ \bibinfo {pages} {091701}
		(\bibinfo {year} {2018})}\BibitemShut {NoStop}%
	\bibitem [{EMS(2018)}]{EMS4}%
	\BibitemOpen
	\href@noop {} {\bibfield  {journal} {\bibinfo  {journal} {Journal of the
				Mechanics and Physics of Solids}\ }\textbf {\bibinfo {volume} {112}},\
		\bibinfo {pages} {577 } (\bibinfo {year} {2018})}\BibitemShut {NoStop}%
	\bibitem [{\citenamefont {Xu}, \citenamefont {Yang},\ and\ \citenamefont
		{Cao}(2018)}]{EMS5}%
	\BibitemOpen
	\bibfield  {author} {\bibinfo {author} {\bibfnamefont {Y.}~\bibnamefont
			{Xu}}, \bibinfo {author} {\bibfnamefont {Z.}~\bibnamefont {Yang}}, \ and\
		\bibinfo {author} {\bibfnamefont {L.}~\bibnamefont {Cao}},\ }\href@noop {}
	{\bibfield  {journal} {\bibinfo  {journal} {Journal of Physics D: Applied
				Physics}\ }\textbf {\bibinfo {volume} {51}},\ \bibinfo {pages} {175106}
		(\bibinfo {year} {2018})}\BibitemShut {NoStop}%
	\bibitem [{\citenamefont {Shen}\ \emph {et~al.}(2018)\citenamefont {Shen},
		\citenamefont {Sun}, \citenamefont {Barnhart},\ and\ \citenamefont
		{Huang}}]{EMS6}%
	\BibitemOpen
	\bibfield  {author} {\bibinfo {author} {\bibfnamefont {X.}~\bibnamefont
			{Shen}}, \bibinfo {author} {\bibfnamefont {C.-T.}\ \bibnamefont {Sun}},
		\bibinfo {author} {\bibfnamefont {M.~V.}\ \bibnamefont {Barnhart}}, \ and\
		\bibinfo {author} {\bibfnamefont {G.}~\bibnamefont {Huang}},\ }\href@noop {}
	{\bibfield  {journal} {\bibinfo  {journal} {Journal of Applied Physics}\
		}\textbf {\bibinfo {volume} {123}},\ \bibinfo {pages} {091708} (\bibinfo
		{year} {2018})}\BibitemShut {NoStop}%
	\bibitem [{\citenamefont {Tol}\ \emph {et~al.}(2017)\citenamefont {Tol},
		\citenamefont {Xia}, \citenamefont {Ruzzene},\ and\ \citenamefont
		{Erturk}}]{EMS7}%
	\BibitemOpen
	\bibfield  {author} {\bibinfo {author} {\bibfnamefont {S.}~\bibnamefont
			{Tol}}, \bibinfo {author} {\bibfnamefont {Y.}~\bibnamefont {Xia}}, \bibinfo
		{author} {\bibfnamefont {M.}~\bibnamefont {Ruzzene}}, \ and\ \bibinfo
		{author} {\bibfnamefont {A.}~\bibnamefont {Erturk}},\ }\href@noop {}
	{\bibfield  {journal} {\bibinfo  {journal} {Applied Physics Letters}\
		}\textbf {\bibinfo {volume} {110}},\ \bibinfo {pages} {163505} (\bibinfo
		{year} {2017})}\BibitemShut {NoStop}%
\end{thebibliography}
%

\end{document}